\documentclass[journal]{IEEEtran}
\usepackage{cite}
\usepackage[pdftex]{graphicx}
\usepackage[cmex10]{mathtools}
\usepackage{amssymb}
\usepackage{epsfig}
\usepackage{subfigure}
\usepackage{algorithm}
\usepackage{algpseudocode}
\usepackage[usenames,dvipsnames]{color}
\usepackage{array}

\usepackage{bm}
\usepackage{bbm}
\usepackage{microtype}
\usepackage{kotex}
\usepackage[T1]{fontenc}
\usepackage[utf8]{inputenc}

\newtheorem{prop}{Proposition}

 \def\bb{{\mathbf{b}}} \def\bc{{\mathbf{c}}} \def\bd{{\mathbf{d}}}
\def\bee{{\mathbf{e}}}  \def\bg{{\mathbf{g}}} \def\bh{{\mathbf{h}}}
   
 \def\bn{{\mathbf{n}}}  
 \def\br{{\mathbf{r}}}  \def\bt{{\mathbf{t}}}
   
 \def\bz{{\mathbf{z}}}  

\def\bA{{\mathbf{A}}}   \def\bD{{\mathbf{D}}}
   
\def\bI{{\mathbf{I}}}  \def\bK{{\mathbf{K}}} 
   \def\bP{{\mathbf{P}}}
\def\bQ{{\mathbf{Q}}}   
   
 \def\bZ{{\mathbf{Z}}}

\DeclareMathOperator*{\argmin}{arg\,min}

\begin{document}


\title{{Joint Vehicle Tracking and RSU Selection for V2I Communications with Extended Kalman Filter}}


\author{Jiho Song,~\IEEEmembership{Member,~IEEE,} Seong-Hwan Hyun,~\IEEEmembership{Student Member,~IEEE,}  Jong-Ho Lee,~\IEEEmembership{Member,~IEEE,}  \\  Jeongsik Choi,~\IEEEmembership{Member,~IEEE,}  and  Seong-Cheol Kim,~\IEEEmembership{Senior Member,~IEEE}
\thanks{This work has been submitted to the IEEE for possible publication. Copyright may be transferred without notice, after which this version may no longer be accessible.}
\thanks{J.\ Song  is with the School of Electrical Engineering, University of Ulsan, Ulsan 44610, South Korea (e-mail: jihosong@ulsan.ac.kr).}
\thanks{S.-H.\ Hyun and S.-C. \ Kim are with the School of Electrical Engineering, Seoul National University, and also with the Institute of New Media \& Communications
(INMC), Seoul 08826, South Korea (e-mail: \{shhyun,   sckim\}@maxwell.snu.ac.kr).}
\thanks{J.-H.\ Lee   \textit{(corresponding author)} is with the School of Electronic Engineering, Soongsil University, Seoul 06978, South Korea (e-mail: jongho.lee@ssu.ac.kr).}
\thanks{J.\ Choi is with the School of Electronics Engineering, Kyungpook National University, Daegu 41566 (e-mail: jeongsik.choi@knu.ac.kr).}}

\maketitle


\begin{abstract}
We develop joint vehicle tracking and road side unit (RSU) selection algorithms suitable for vehicle-to-infrastructure (V2I) communications.
We first  design an analytical framework for evaluating  vehicle tracking systems based on the extended Kalman filter.
A simple, yet effective, metric  that  quantifies  the vehicle tracking performance is derived in terms of the angular derivative of a dominant spatial frequency.
Second, an RSU selection algorithm is proposed to select a proper RSU that enhances the vehicle tracking performance.
{A joint vehicle tracking algorithm is also developed to maximize the tracking performance by  considering sounding samples at multiple  RSUs while minimizing the amount of sample exchange.}
The numerical results  verify that the proposed vehicle tracking algorithms give better performance than conventional signal-to-noise ratio-based tracking systems.
\end{abstract}


\IEEEpeerreviewmaketitle

\section{Introduction}

Vehicle-to-everything (V2X) communications are a promising candidate to facilitate  intelligent transport systems  required for fully connected vehicular networks  \cite{Ref_Zhang11,Ref_TS_22_186}.
Vehicles can obtain traffic information beyond the sensing range of radar and LIDAR via  wireless communication networks \cite{Ref_Chg16Mag}.
{V2X-assisted collaborate sensing are thought to be one of the key enabling technologies in fully automated driving \cite{Ref_Kut19}.}
The enhanced traffic situational awareness via a wireless network would create new services, e.g., vehicular platoon driving \cite{Ref_Lee20}.
{Vehicular communication networks are a prime technology for supporting intelligent transportation system.}

Securing a seamless radio connection is a significant challenge in vehicle-to-infrastructure (V2I) communication networks.
Such connections are required to utilize a wide bandwidth in the millimeter-wave (mmWave) spectrum because a large amount of sensor data must be exchanged  \cite{Ref_Chg16Mag}.
The vehicular channels at mmWave frequencies may change rapidly because the relative velocity between vehicles and a mounted road side unit (RSU) becomes large compared to conventional mobile systems.
{Unpredictable vehicle movements make it difficult to estimate future trajectories of fast-moving vehicles \cite{Ref_Nog20}.
Fast and reliable vehicle tracking  is essential to  cope with  channel fluctuations \cite{Ref_Zen19}.
The  challenge in V2I communications is to maintain wireless connections while satisfying these conflicting requirements for vehicle tracking \cite{Ref_Va16}.}

Assuming a single cell network, vehicle tracking algorithms, which can be incorporated in current cellular networks, have been developed based on the extended Kalman filter (EKF) \cite{Ref_Sha19,Ref_Hyu22}.
In V2I networks, RSUs must be densely deployed to avoid an intermittent disconnection of the radio link \cite{Ref_Bar13}.
A handover between RSUs frequently occurs due to the high mobility characteristics of vehicular channels \cite{Ref_Cho18_handover}.
The road environment with densely deployed RSUs necessitates the development of a fast and reliable RSU selection algorithm to secure a seamless radio connection.
Considering a multi-connectivity framework, a joint vehicle tracking algorithm is also needed to maximize the vehicle tracking performance.

{Coordinated multi-point transmission algorithms} have been studied extensively to enhance network throughput {\cite{Ref_Irm11}.
Conventional multi-transmission systems are designed based on an average signal-to-noise ratio (SNR) metric that is suitable for evaluating cellular networks with a circular shape  service area.
On the other hand, the  shape of the service area in V2I networks is very different from cellular networks because most roads are very narrow {and RSUs are installed near the road \cite{Ref_Hyu22}.}
Therefore, the SNR-based transmission systems {may be unsuitable} for V2I networks.
Furthermore, the tight budget constraints of RSUs necessitate the redesign of communication systems  cost-effectively to enable low latency communications using low-cost transceivers.

{In this paper, we aim to develop an analytical framework of joint vehicle tracking and RSU selection for V2I communications.
First, we design a simple metric to quantify a vehicle tracking performance accurately by considering the unique shape of a service area in V2I networks.
To the best of the author’s knowledge,  a vehicle tracking performance metric has not yet been designed in terms of the angular variation in a spatial frequency domain.}
{Second, based on the proposed metric,} an RSU selection algorithm is developed for providing a seamless handover between neighboring RSUs.
{Lastly, a joint vehicle tracking algorithm is developed to maximize   tracking performance by considering  angular variations obtained from multiple RSUs while minimizing the amount of data exchange.}

\section{System Model}
\label{sec:SM}

We consider a vehicle tracking system  designed  based on the EKF with uplink channel sounding samples.
Considering a multiple-input single-output system using $M$ transmit antennas, an input-output expression for the $\ell$-th received sounding sample  is defined by
\begin{align*}
r_{\ell}^{u}=\sqrt{\rho_{\ell}^u} \bz_{\ell}^u\bh_{\ell}^{u}+  n_{\ell}^{u},
\end{align*}
where $u \in \{1,\cdots, U\}$ is the index for the RSU,  {$r_{\ell}^{u} = r_{\ell}^{u,\textrm{re}}+j r_{\ell}^{u,\textrm{im}} \in \mathbb{C}$ is the received sounding sample, $\bz_{\ell}^u= \bz_{\ell}^{u,\textrm{re}}+j \bz_{\ell}^{u,\textrm{im}}  \in \mathbb{C}^{1 \times M}$} is the unit-norm combing vector at the RSU, {$\bh_{\ell}^{u} = {\bh}_{\ell}^{u,\textrm{re}} +j {\bh}_{\ell}^{u,\textrm{im}}  \in \mathbb{C}^{M}$} is the channel vector, and {$n_{\ell}^{u} = n_{\ell}^{u,\textrm{re}} + j n_{\ell}^{u,\textrm{im}}\sim \mathcal{CN}(0,1)$} is the normalized (combined) noise that follows complex Gaussian distribution with zero mean and unit variance.
The  average SNR is $\rho_{\ell}^u \doteq \frac{\varrho \mathrm{G}_{\mathrm{tx}}  \mathrm{G}_{\mathrm{rx}} }{\sigma_{\mathrm{n}}^2}  \big(\frac{\lambda}{4\pi {d}_{\ell}^u}\big)^n \stackrel{(a)} = \frac{\varrho}{\sigma_{\mathrm{n}}^2} \big(\frac{\lambda}{4\pi {d}_{\ell}^u}\big)^n$, where $\varrho$ is the transmit power, $\sigma_{\mathrm{n}}^2$ is the noise power, ${d}_{\ell}^u$ is the distance between a vehicle and the $u$-th RSU, $\lambda$ is the wavelength of the radio signals, and $n$ is the path-loss exponent.
Note that $\mathrm{G}_{\mathrm{tx}} = \mathrm{G}_{\mathrm{rx}} = 1$ is assumed in $(a)$ to simplify {the} presentation.
By decomposing all the variables into real and imaginary parts, the input-output expression is rewritten in {the} real domain, as
\begin{align}
\label{eq:sounding_real}
\tilde{\br}_{\ell}^{u}= \sqrt{\rho_{\ell}^u}\tilde{\bZ}_{\ell}^{u} \tilde{\bh}_{\ell}^{u} + \tilde{\bn}_{\ell}^{u},
\end{align}
where $\tilde{\br}_{\ell}^{u}=[r_{\ell}^{u,\textrm{re}}, r_{\ell}^{u,\textrm{im}}]^T \in \mathbb{R}^{2}$, $\tilde{\bZ}_{\ell}^{u} = \Big[
             \begin{array}{cc}
               \bz_{\ell}^{u,\textrm{re}} & -\bz_{\ell}^{u,\textrm{im}} \\
               \bz_{\ell}^{u,\textrm{im}} & \bz_{\ell}^{u,\textrm{re}} \\
             \end{array}
\Big] \in \mathbb{R}^{2 \times 2M} $, $\tilde{\bh}_{\ell}^{u}=[({\bh}_{\ell}^{u,\textrm{re}})^T, ({\bh}_{\ell}^{u,\textrm{im}})^T]^T \in \mathbb{R}^{2M}$, and $\tilde{\bn}_{\ell}^{u} =[n_{\ell}^{u,\textrm{re}}, n_{\ell}^{u,\textrm{im}}]^T \in \mathbb{R}^{2}$.

\begin{figure}
\centering
\subfigure{\includegraphics[width=0.375\textwidth]{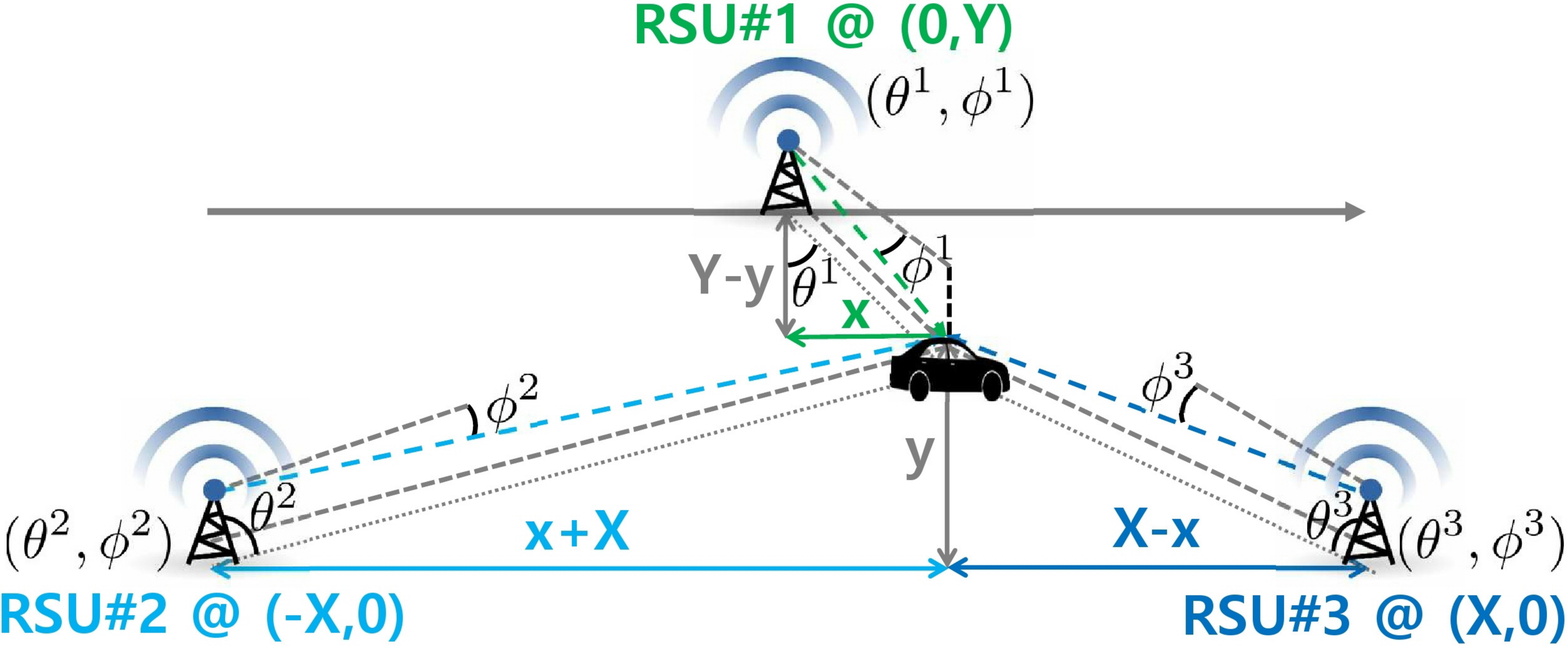}}
\caption{An overview of V2I communication system.}
\label{fig:system_overview}
\end{figure}

{By assuming} a line-of-sight channel with a dominant radio path, the channel vector can be modeled by
\begin{align*}
\bh_{\ell}^{u} \simeq \bh(\psi_{\ell}^{u}) \doteq \beta_{\ell}^{u} \bd_{M}(\psi_{\ell}^{u}),
\end{align*}
where $\beta_{\ell}^{u}$ is the small-scale channel fading parameter, {and $\bd_{M}(\psi_{\ell}^{u}) =[1,e^{j\psi_{\ell}^{u}},\cdots,e^{j(M-1)\psi_{\ell}^{u}}]^T \in \mathbb{C}^{M}$ is an array response vector with a spatial frequency, $\psi_{\ell}^{u} \in [-\pi, \pi)$.}
The real and imaginary parts of the array response vector are {expressed as}
\begin{align*}
\tilde{\bd}_{M}^{\textrm{re}}(\psi_{\ell}^{u})&=\big[\cos(0),\cos(\psi_{\ell}^{u}),\cdots, \cos((M-1)\psi_{\ell}^{u})\big]^T \in \mathbb{R}^{M},
\\
\tilde{\bd}_{M}^{\textrm{im}}(\psi_{\ell}^{u})&=\big[\sin(0),\sin(\psi_{\ell}^{u}),\cdots, \sin((M-1)\psi_{\ell}^{u})\big]^T \in \mathbb{R}^{M}.
\end{align*}
The channel vector is then reformulated in {a} real domain as
\begin{align*}
\tilde{\bh}_{\ell}^{u} \simeq \tilde{\bh}(\psi_{\ell}^{u})=
\bigg[\begin{array}{cc}
               \beta_{\ell}^{u,\textrm{re}} \tilde{\bd}_{M}^{\textrm{re}}(\psi_{\ell}^{u}) - \beta_{\ell}^{u,\textrm{im}}\tilde{\bd}_{M}^{\textrm{im}}(\psi_{\ell}^{u})  \\
              \beta_{\ell}^{u,\textrm{im}} \tilde{\bd}_{M}^{\textrm{re}}(\psi_{\ell}^{u}) + \beta_{\ell}^{u,\textrm{re}}\tilde{\bd}_{M}^{\textrm{im}}(\psi_{\ell}^{u})  \\
             \end{array}\bigg] \in \mathbb{R}^{2M}.
\end{align*}

{Fig. \ref{fig:system_overview} presents the RSU deployment scenario, where $h$ is the relative height difference between the RSU and vehicles.}
The position on the $y$-axis is static because {it is assumed that} the vehicle does not change the traffic lane.
The spatial frequencies for the RSUs can be written by
\begin{align*}
\psi_{\ell}^{1} &= \pi \sin\theta_{\ell}^1 \cos \phi_{\ell}^1=\frac{\pi x_{\ell}}{(x_{\ell}^2+(\mathrm{Y}-y)^2+h^2)^{\frac{1}{2}}} \doteq \mathrm{g}^{1}(\bt_{\ell}),
\\
\psi_{\ell}^{2} &= \pi \sin\theta_{\ell}^2 \cos \phi_{\ell}^2= \frac{\pi (\mathrm{X}+x_{\ell})}{ ((\mathrm{X}+x_{\ell})^2+y^2+h^2)^{\frac{1}{2}}} \doteq \mathrm{g}^{2}(\bt_{\ell}),
\\
\psi_{\ell}^{3} &= \pi \sin\theta_{\ell}^3 \cos \phi_{\ell}^3=\frac{\pi ({\mathrm{X}-x_{\ell}})}{(({\mathrm{X}-x_{\ell}})^2+y^2+h^2)^{\frac{1}{2}}} \doteq \mathrm{g}^{3}(\bt_{\ell}),
\end{align*}
because the horizontal and vertical angle of departure (AoD) for the $u$-th RSU, i.e., $\theta_{\ell}^u$ and $\phi_{\ell}^u$, are defined as functions of {the}  position variables, $(x_{\ell},y)$, and network  parameters, $(\mathrm{X},\mathrm{Y},h)$.

The vehicle movement can be {modeled} by using the linear state transition model with a sampling period $T_s$, such as
\begin{align}
\label{eq:predict}
\bt_{\ell}= \bA \bt_{\ell-1} + \bb \alpha + \bc_{\ell-1},
\end{align}
where  $\bt_{\ell}=[x_{\ell}, v_{\ell}]^T \in \mathbb{R}^2$ is the state vector of a vehicle, in which $x_{\ell}$ and $v_{\ell}$ denote a position on the $x$-axis and {the} velocity of the vehicle at the $\ell$-th channel use, and $\alpha \sim \mathcal{N}(0,\sigma_{\alpha}^2)$ is the acceleration parameter that follows Gaussian distribution with zero mean and variance $\sigma_{\alpha}^2$  \cite{Ref_Hyu22}.
A state transition matrix, an acceleration transition vector, and an error transition vector are defined {as,}
$\bA=\bigg[
             \begin{array}{cc}
               1 & T_s \\
               0 & 1 \\
             \end{array}
           \bigg]
 \in \mathbb{R}^{2 \times 2}$,  $\bb=\Big[\frac{T_s^2}{2}, T_s\Big]^T \in \mathbb{R}^{2}$, and $\bc_{\ell-1}  \sim \mathcal{N}(\mathbf{0}_2, \bQ_{\omega})$, respectively \cite{Ref_Ken18}.
Similar to \cite{Ref_Sha19}, {the} covariance matrices for the acceleration transition vector and the error transition vector are  modeled by
$\bQ_{\alpha}=\bb\bb^T \sigma_{\alpha}^2 \in \mathbb{R}^{2 \times 2}$ and $\bQ_{\omega}=\sigma_{\omega}^2\mathrm{diag}[T_s^2, 1 ] \in \mathbb{R}^{2 \times 2}$, respectively.
{It is assumed} that $\bc_{\ell-1}$ is dynamic, while $\alpha$ is static in  each coherence time.

{This paper reviews} a vehicle tracking algorithm developed based on {the} EKF \cite{Ref_Hyu22}.
Assuming the $u$-th RSU is used for vehicle tracking, {the} initial state information of {a} vehicle, $(x_0,y,v_0)$, is fed back to the RSU.
Furthermore,  {it is assumed} that  channel fading information, $(\rho_{\ell}^{u},\beta_{\ell}^{u})$, is  known at the RSU.
\subsubsection{State prediction process}
{Since RSU does not have information on the acceleration and error transition vectors, the vehicle state vector is} predicted based on the state transition model {as follows:}
\begin{align}
\label{eq:state_predict}
\hat{\bt}_{\ell|\ell-1}^{u}=\bA\hat{\bt}_{\ell-1}^{u},
\end{align}
where $\hat{\bt}_{\ell-1}^{u}$ is the state vector estimated at discrete time $\ell-1$.
{Without correcting the transition errors, the covariance matrix of the prediction error will be updated as} $\hat{\bQ}_{\ell|\ell-1}^{u}=\bA \hat{\bQ}_{\ell-1}^{u} \bA^T + \bQ_{\mathrm{e}}$, where $\hat{\bQ}_{\ell-1}^{u}$ is {the} estimated covariance matrix at {a} discrete time $\ell-1$, and $\bQ_{\mathrm{e}}=\bQ_{\alpha}+\bQ_{\omega}$.

\begin{figure*}
\centering
\subfigure[Proposed SANR-based RSU selection system.]{\label{fig:hand_over_example_01}\includegraphics[width=0.4125\textwidth]{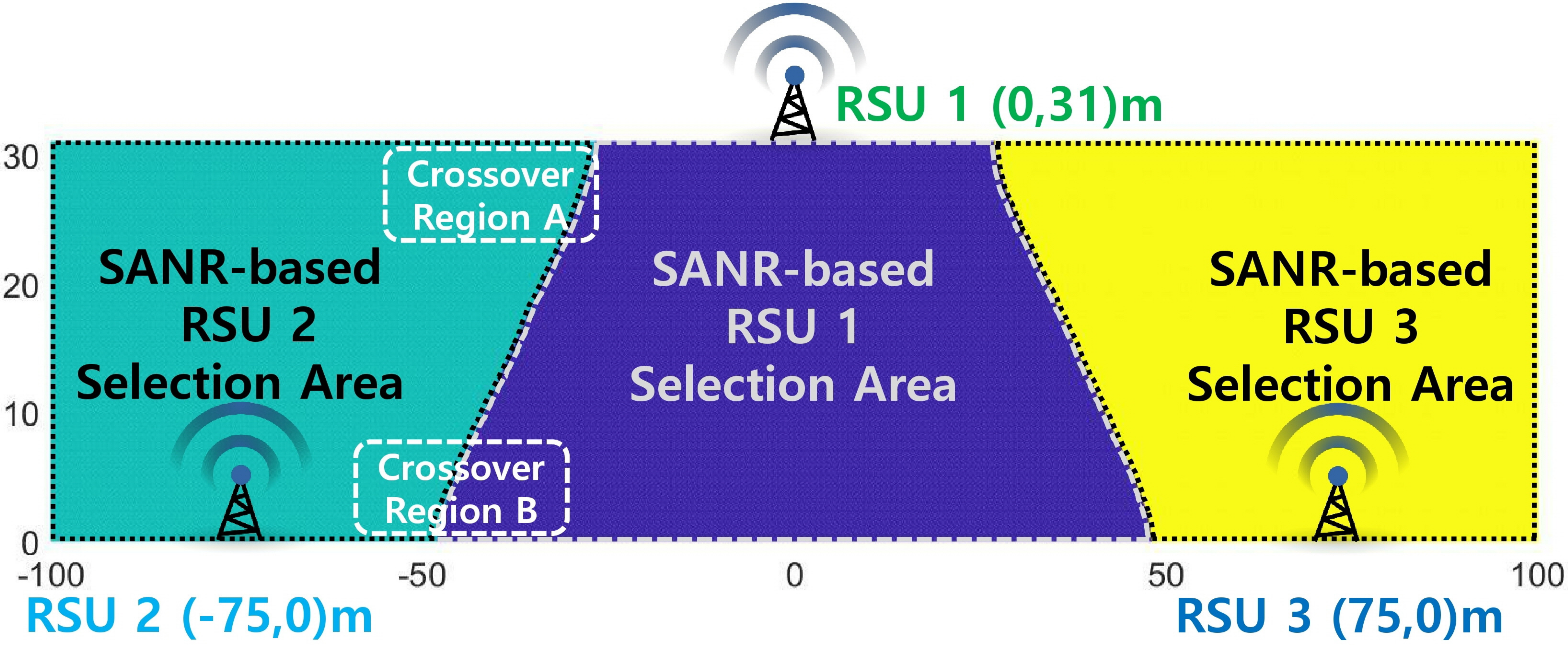}}
\subfigure[SNR-based RSU selection system.]{\label{fig:hand_over_example_02}\includegraphics[width=0.4125\textwidth]{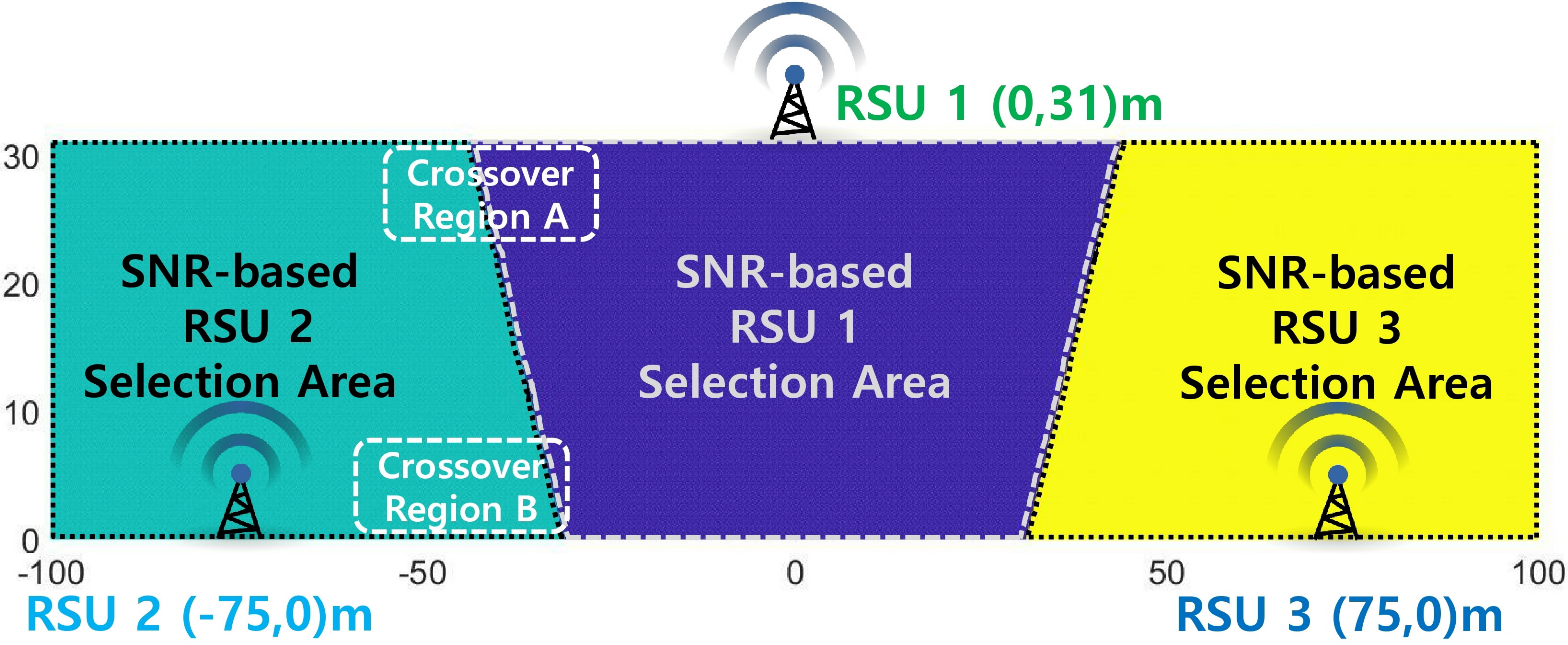}}
\subfigure[{Proposed SANR-based joint tracking system with $\tau_{\textrm{th}}=0.98$.}]{\label{fig:hand_over_example_03}\includegraphics[width=0.4125\textwidth]{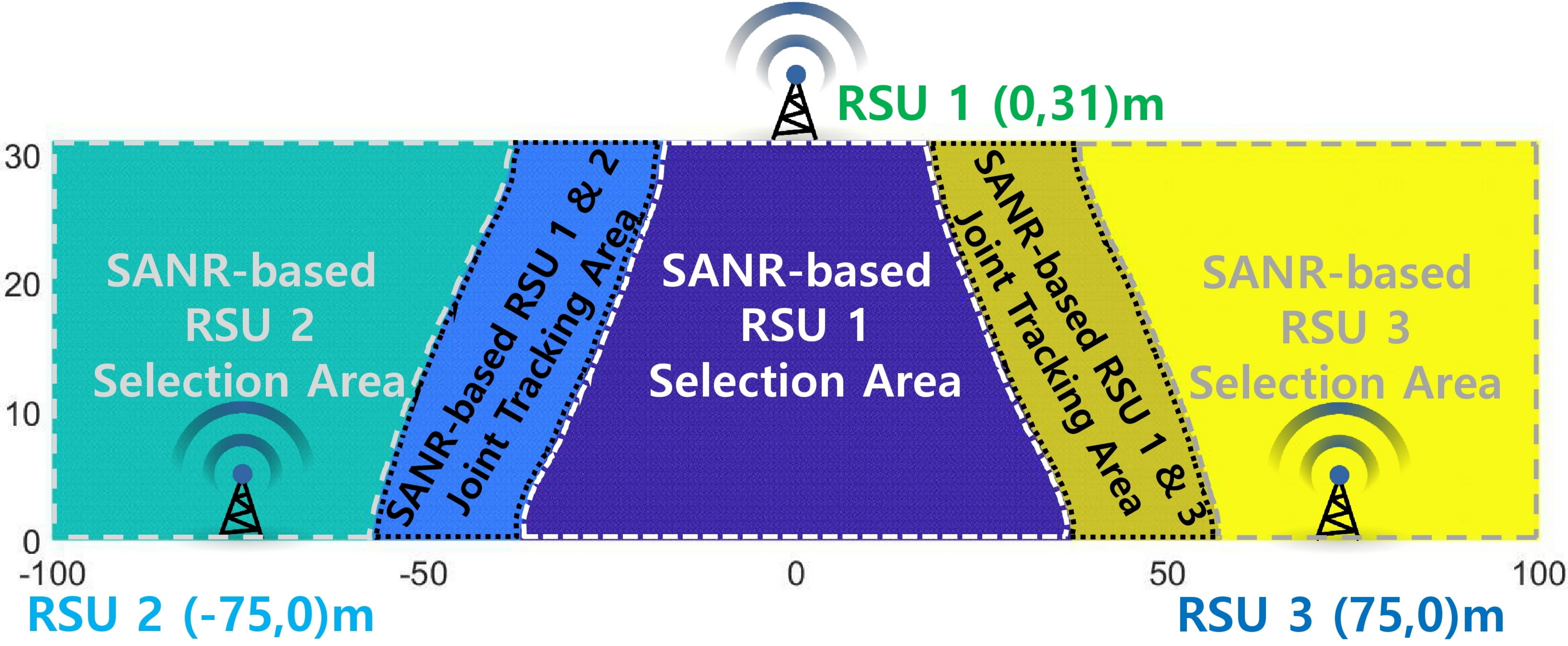}}
\subfigure[{SNR-based joint tracking system  with $\tau_{\textrm{th}}=0.662$.}]{\label{fig:hand_over_example_04}\includegraphics[width=0.4125\textwidth]{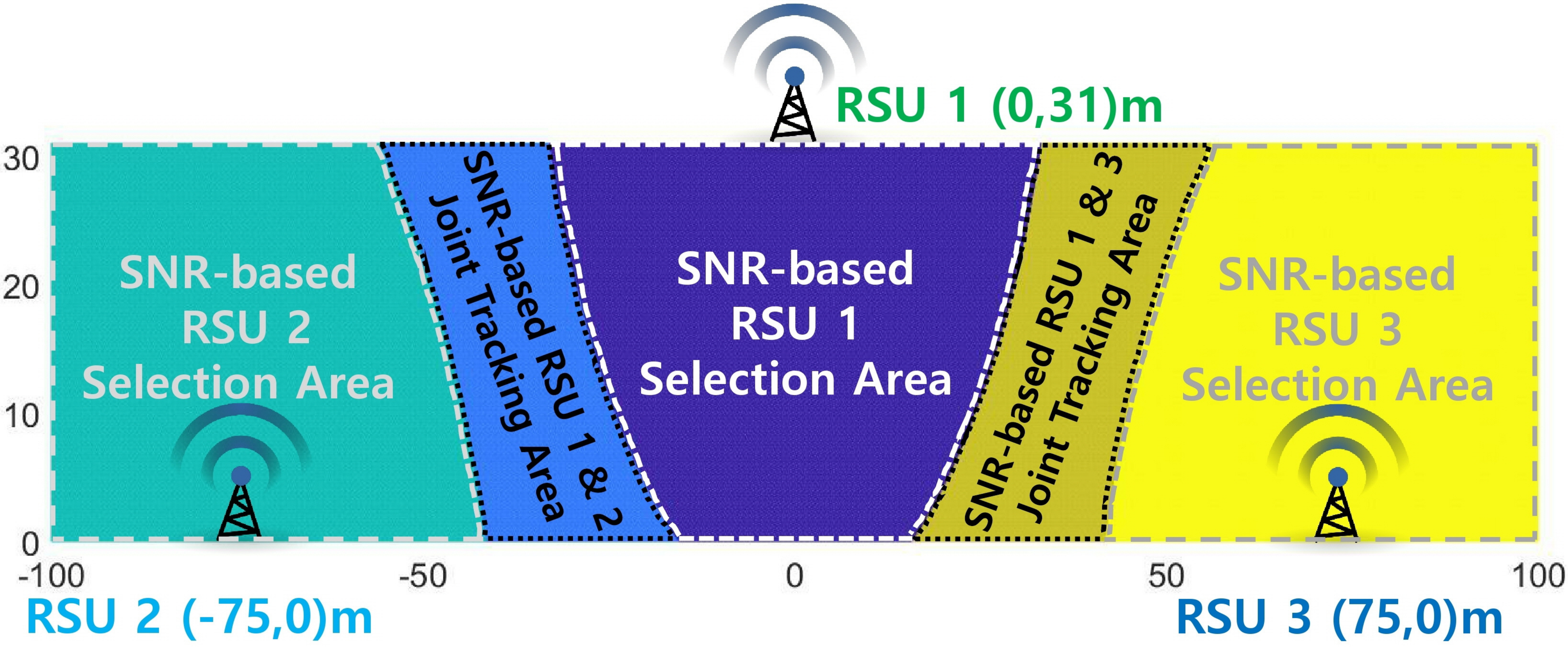}}
\caption{{Service areas for EKF-based vehicle tracking systems.}}
\label{fig:hand_over_example}
\end{figure*}

\subsubsection{State update process} The predicted state vector will be refined   using the channel sounding sample in (\ref{eq:sounding_real}), such as
\begin{align}
\label{eq:state_update}
\hat{\bt}_{\ell}^{u}={\hat{\bt}_{\ell|\ell-1}}^{u} +\tilde{\bK}_{\ell}^{u}\big(\tilde{\br}_{\ell}^{u}-\sqrt{\rho_{\ell}^u} \tilde{\bZ}_{\ell}^{u} \tilde{\bh}(\hat{\psi}_{\ell|\ell-1}^{u}) \big),
\end{align}
where $\tilde{\bh}(\hat{\psi}_{\ell|\ell-1}^{u})$ is the ray-like channel that is predicted  by using the estimated spatial frequency, $\hat{\psi}_{\ell|\ell-1}^{u}=\mathrm{g}^{u}(\hat{\bt}_{\ell|\ell-1}^{u})$.
{The combiner at  RSU is designed to minimize the trace of the estimated covariance matrix, such that $\tilde{\bZ}_{\ell}^{u} = \argmin \mathrm{Tr} \big\{ \hat{\bQ}_{\ell}^{u} \big\}$.
Please refer to the combiner design process in \cite{Ref_Hyu22} for $\tilde{\bZ}_{\ell}^{u}$.}
The state estimation error is refined as
\begin{align*}
\hat{\bQ}_{\ell}^{u}=\big(\bI_2- \sqrt{\rho_{\ell}^u} \tilde{\bK}_{\ell}^{u} \tilde{\bZ}_{\ell}^{u} \tilde{\bD}_{\ell|\ell-1}^{u}\big)\hat{\bQ}_{\ell|\ell-1}^{u},
\end{align*}
where the Kalman gain matrix is
\begin{align}
\nonumber
\tilde{\bK}_{\ell}^{u}&=\sqrt{\rho_{\ell}^u}\hat{\bQ}_{\ell|\ell-1}^{u}   ( \tilde{\bZ}_{\ell}^{u} \tilde{\bD}_{\ell|\ell-1}^{u} )^T
\\
\label{eq:kalman}
&~~~~~~~\big(\rho_{\ell}^u \tilde{\bZ}_{\ell}^{u} \tilde{\bD}_{\ell|\ell-1}^{u} \hat{\bQ}_{\ell|\ell-1}^{u}  (\tilde{\bZ}_{\ell}^{u} \tilde{\bD}_{\ell|\ell-1}^{u} )^T +{\bI_2}/2  \big)^{-1}.
\end{align}
The Jacobian matrix of the channel  is defined by $\tilde{\bD}_{\ell|\ell-1}^{u} = \dot{\bh}^{u}_{\ell|\ell-1}(\dot{\bg}^{u}_{\ell|\ell-1})^T$, where $\dot{\bh}^{u}_{\ell|\ell-1} = \frac{\partial \tilde{\bh}(\psi)}{\partial \psi}\Big|_{\psi=\mathrm{g}^{u}(\hat{\bt}_{\ell|\ell-1}^{u})}$ is the partial derivative of the channel, and {the partial derivative of the spatial frequency for the RSUs can be computed as
\begin{align}
\label{eq:pd_spa_freq}
\dot{\bg}_{\ell|\ell-1}^{u}&= \frac{\partial  \mathrm{g}^{u}(\bt) }{\partial \bt}\bigg|_{\bt=\hat{\bt}_{\ell|\ell-1}^{u}}=\pi \dot{\mathrm{g}}_{\ell|\ell-1}^{u}[1, T_s]^T ,
\\
\nonumber
\dot{\mathrm{g}}_{\ell|\ell-1}^{1} & = ((\mathrm{Y}-y)^2+h^2)(\hat{x}_{\ell|\ell-1}^2+(\mathrm{Y}-y)^2+h^2)^{-\frac{3}{2}},
\\
\nonumber
\dot{\mathrm{g}}_{\ell|\ell-1}^{2} & = (y^2+h^2)((\mathrm{X}+\hat{x}_{\ell|\ell-1})^2+y^2+h^2)^{-\frac{3}{2}},
\\
\nonumber
\dot{\mathrm{g}}_{\ell|\ell-1}^{3} & = {-}(y^2+h^2)(({\mathrm{X}-\hat{x}_{\ell|\ell-1}})^2+y^2+h^2)^{-\frac{3}{2}}.
\end{align}
Please refer to  \cite{Ref_Hyu22} to calculate the Jacobian matrix.}

\section{RSU Selection Algorithm for V2I Handover}
\label{sec:hand_over_algorithm}

{An RSU selection algorithm aims to choose an RSU that can provide the best vehicle tracking performance.}
In cellular networks, mobile users switch the base station based on the average SNR.
Similar to cellular networks, one possible method of RSU selection  is to {select an RSU} that can provide the largest  SNR,   $u =\max_{u \in \{1,\cdots, U\}}\rho^{u}_{\ell}$.
With the largest  SNR, the service region for RSUs can be defined, {as shown in Fig. \ref{fig:hand_over_example_02}.}

{This paper takes} a closer look at the state update process to evaluate the EKF-based vehicle tracking systems analytically.
The state update process is designed to correct the state transition error, ${\bee}_{\ell|\ell-1}^{u} \doteq \bt_{\ell}-\hat{\bt}_{\ell|\ell-1}^{u}$,  {using  channel-sounding samples.}
In the EKF-based vehicle tracking algorithm, {the} dominant spatial frequency  obtained from a sounding sample will be used to correct the state transition error.
The state update process becomes immune to noise as the  angular {variations caused by vehicle movements become larger.}
Angular variation might be significantly different depending on the relative location between a vehicle and RSUs, {even though the} vehicle travels the equivalent distances.

{In the state update process, additive} noise should be compared {with the angular variation} instead of evaluating the signal strength of the sounding sample using an average SNR.
{A new metric  is   derived to  quantify a vehicle tracking performance in terms of an angular variation.}
In (\ref{eq:state_update}), the  sounding sample for the $u$-th RSU is compared with the predicted   sample, such as
\begin{align}
\nonumber
\Gamma_{\ell}^{u}&=\tilde{\br}_{\ell}^{u}-\sqrt{\rho_{\ell}^u}\tilde{\bZ}_{\ell}^{u}\tilde{\bh}(\hat{\psi}_{\ell|\ell-1}^{u})
\\
\nonumber
&=\sqrt{\rho_{\ell}^u}\tilde{\bZ}_{\ell}^{u}(\tilde{\bh}_{\ell}^{u}-\tilde{\bh}(\hat{\psi}_{\ell|\ell-1}^{u})) +\tilde{\bn}_{\ell}^{u}
\\
\label{eq:Gamma}
& \stackrel{(a)} \simeq \sqrt{\rho_{\ell}^u}\underbrace{\tilde{\bZ}_{\ell}^{u} \tilde{\bD}_{\ell|\ell-1}^{u}{\bee}^u_{\ell|\ell-1}}_{(b)}+\tilde{\bn}_{\ell}^{u},
\end{align}
where $(a)$ is derived using the approximated channel vector $\tilde{\bh}_{\ell}^{u} \simeq \tilde{\bh}(\hat{\psi}_{\ell|\ell-1}^{u}) + \tilde{\bD}_{\ell|\ell-1}^{u}{\bee}^u_{\ell|\ell-1}$, and the predicted channel, $\tilde{\bh}(\hat{\psi}_{\ell|\ell-1}^{u})$,  is defined by using the estimated spatial frequency, $\hat{\psi}_{\ell|\ell-1}^{u}=\mathrm{g}^{u}(\hat{\bt}_{\ell|\ell-1}^{u})$.
The vehicle estimation performance   depends on the power ratio between the state-error-correction component\footnote{The state-error-correction component denotes the correlation between the state transition error and the combined Jacobian matrix that includes the derivative of a spatial frequency with respect to a change in the state vector.} in $(b)$ and the noise  component.
In the following proposition,   {signal-plus-angular-derivative-to-noise ratio (SANR)} {is defined as a function of the position variables.}

\begin{prop}
\label{prop:01}
The SANRs of  sounding samples, {$\gamma^{u}_{\ell|\ell-1}$ with $u \in \{1,2,3\}$, are approximated by}
\begin{align*}
\gamma_{\ell|\ell-1}^{1} &\simeq {\kappa\big((\mathrm{Y}-y)^2+h^2\big)^2}{\big(\hat{x}_{\ell|\ell-1}^2+(\mathrm{Y}-y)^2+h^2\big)^{-(3+\frac{n}{2})}},
\\
\gamma_{\ell|\ell-1}^{2} &\simeq {\kappa\big(y^2+h^2\big)^2}{\big((\mathrm{X}+\hat{x}_{\ell|\ell-1})^2+y^2+h^2\big)^{-(3+\frac{n}{2})}},
\\
\gamma_{\ell|\ell-1}^{3} &\simeq {\kappa\big(y^2+h^2\big)^2}{\big(({\mathrm{X}-\hat{x}_{\ell|\ell-1}})^2+y^2+h^2\big)^{-(3+\frac{n}{2})}},
\end{align*}
where $\kappa=\frac{\varrho(M-1)(2M-1)\pi^2\lambda^n  {(\hat{\bQ}_{\ell|\ell-1})_{1,1}} }{6(4\pi)^n \sigma_{\mathrm{n}}^2}$.
\end{prop}
\begin{IEEEproof}
The SANR of a sounding sample is derived as,
\begin{align}
\label{eq:eff_snr}
&\gamma_{\ell|\ell-1}^{u}=\rho^{u}_{\ell} \mathrm{E}\big[ \|  \tilde{\bZ}_{\ell}^{u} \tilde{\bD}_{\ell|\ell-1}^{u}  \bee^u_{\ell|\ell-1} \|_2^2  \big]
\\
\nonumber
& \stackrel{(a)}= \frac{\rho^{u}_{\ell}}{M} \mathrm{E}\big[ (\bee^u_{\ell|\ell-1})^T  \dot{\bg}_{\ell|\ell-1}^{u} (\dot{\bh}_{\ell|\ell-1}^{u})^T\dot{\bh}_{\ell|\ell-1}^{u} (\dot{\bg}^{u}_{\ell|\ell-1})^T     \bee^u_{\ell|\ell-1}   \big]
\\
\nonumber
& \stackrel{(b)}= \frac{\rho^{u}_{\ell}(M-1)(2M-1)\mathrm{E}[ (\bee_{\ell|\ell-1}^u)^T  \dot{\bg}_{\ell|\ell-1}^{u}  (\dot{\bg}^{u}_{\ell|\ell-1})^T     \bee_{\ell|\ell-1}^u ]}{6},
\end{align}
where $(a)$ is because $(\tilde{\bZ}_{\ell}^{u})^T \tilde{\bZ}_{\ell}^{u}=\frac{\bI_{2M}}{M}$,  and $(b)$ is derived with
\begin{align*}
&\mathrm{E}\big[(\dot{\bh}_{\ell|\ell-1}^{u})^T\dot{\bh}_{\ell|\ell-1}^{u}\big] = {\mathrm{E}\big[|\beta^{u,\mathrm{re}}_{\ell}|^2+|\beta^{u,\mathrm{im}}_{\ell}|^2 \big] \| \dot{\bd}_{M} ( \hat{\psi}_{\ell|\ell-1}^{u} ) \|_2^2}
\\
&=\sum_{m=0}^{M-1}m^2(\sin^2(m \hat{\psi}_{\ell|\ell-1}^{u})+\cos^2(m \hat{\psi}_{\ell|\ell-1}^{u}))
\\
&=\frac{M(M-1)(2M-1)}{6}.
\end{align*}
The SANR in $(\ref{eq:eff_snr})$ is approximated by considering $T_s \ll 1$, {as}
\begin{align*}
\gamma_{\ell|\ell-1}^{u} &\stackrel{(a)} \simeq  \frac{\rho_{\ell}^u(M-1)(2M-1) \pi^2   (\dot{\mathrm{g}}_{\ell|\ell-1}^{u})^2\mathrm{E}[|(\bee^u_{\ell|\ell-1})_{1,1}|^2] }{6}
\\
&\stackrel{(b)}= \frac{\varrho(M-1)(2M-1)  \pi^2  \lambda^n (\dot{\mathrm{g}}_{\ell|\ell-1}^{u})^2{(\hat{\bQ}^{u}_{\ell|\ell-1})_{1,1}}}{6(4\pi {d}_{\ell}^u)^n \sigma_{\mathrm{n}}^2  },
\end{align*}
where {the partial derivative of the spatial frequency, $\dot{\bg}_{\ell|\ell-1}^{u}$, is derived in (\ref{eq:pd_spa_freq})}, the approximation in $(a)$ is derived {from}
\begin{align*}
\dot{\bg}_{\ell|\ell-1}^{u}    (\dot{\bg}^{u}_{\ell|\ell-1})^T              & \simeq\bigg[\begin{array}{cc}
               (\pi \dot{\mathrm{g}}_{\ell|\ell-1}^{u})^2  & 0 \\
              0 &0 \\
             \end{array}\bigg],
\end{align*}
 $(b)$ is based on $\rho_{\ell}^u =\frac{\varrho}{\sigma_{\mathrm{n}}^2} \big(\frac{\lambda}{4\pi {d}_{\ell}^u}\big)^n$, and $\mathrm{E}[|(\bee^u_{\ell|\ell-1})_{1,1}|^2]=(\hat{\bQ}^{u}_{\ell|\ell-1})_{1,1}$.
Note that $(\bA)_{a,b}$ is the $(a,b)$-th entry of  $\bA$.
It is assumed that the first element of the covariance matrix are the same, $(\hat{\bQ}_{\ell|\ell-1})_{1,1}=(\hat{\bQ}^{u}_{\ell|\ell-1})_{1,1}$ for all $u \in \{1,\cdots, U\}$.
\end{IEEEproof}

Assuming a covariance matrix  has arbitrary, but fixed values, the service region  is predefined based on SANRs that can be computed  using vehicle position variables.
For a given predicted state vector,  an RSU generating the largest SANR can be chosen based on the predefined service region in Fig. \ref{fig:hand_over_example_01}.
For example,  vehicles around \textit{Crossover Region A} in Fig. \ref{fig:hand_over_example} will be connected to  RSU $2$ based on the predefined service area, even though the vehicles are close to RSU $1$.
When handover between {the RSUs} is required, a serving RSU will inform {the next RSU of the need for a  handover process.}

\section{Joint Vehicle Tracking Algorithm}
\label{sec:joint_beam_tracking_algorithm}

\begin{algorithm}
  \caption{Proposed SANR-based joint vehicle tracking}
  \label{Al:01}
  \begin{algorithmic}
\State \textbf{{Initialization}}
\State ~1:~~{Initial state vector},~$\hat{\bt}_{0}=[x_0, v_0]^T$
\State ~2:~~{Initial covariance matrix},~$\hat{\bQ}_{0}=\mathbf{0}_{2 \times 2}$
\State \textbf{{Linear state prediction}}
\State ~3:~~{Predict state vector},~$\hat{\bt}_{\ell|\ell-1}= \bA \hat{\bt}_{\ell-1}$
\State ~4:~~{Predict covariance matrix},~$\hat{\bQ}_{\ell|\ell-1}=\bA \hat{\bQ}_{\ell-1} \bA^T + \bQ_{\mathrm{e}}$
\State \textbf{{RSU selection  for joint vehicle tracking}}
\State ~5:~~{Select a set of RSUs, $\mathcal{U}$, based on SANR }
\State ~~~~~${\gamma}^{u_1} \geq {\gamma}^{u_2} \geq {\gamma}^{u_3}$ with $\bar{\gamma} \doteq \sum_{u}\gamma^{u}$
\State ~6:~~{if $\frac{{\gamma}^{{u}_{1}}}{\bar{\gamma}} \geq \tau_{\textrm{th}}$, then $\mathcal{U}=\{ u_1 \}$}
\State ~7:~~{elseif $\frac{{\gamma}^{{u}_{1}}}{\bar{\gamma}} < \tau_{\textrm{th}}$ and $\frac{{\gamma}^{{u}_{1}}+{\gamma}^{{u}_{2}}}{\bar{\gamma}} \geq \tau_{\textrm{th}}$, then $\mathcal{U}=\{ {u}_1,{u}_2 \}$}
\State ~8:~~{else $\mathcal{U}=\{1,2,3 \}$}
\State \textbf{{Uplink channel sounding}}
\State ~9:~~{Selected RSUs compute combiner for sounding, $\tilde{\bZ}^{u}_{\ell}$}
\State 10:~~{Conduct  channel sounding},~$\tilde{\br}^{u}_{\ell}= \sqrt{\rho_{\ell}^u}\tilde{\bZ}^{u}_{\ell} \tilde{\bh}^{u}_{\ell} + \tilde{\bn}^{u}_{\ell}$
\State 11:~~{Exchange samples, $\tilde{\br}_{\ell}^{u}$, between selected RSUs, $u \in \mathcal{U}$}
\State 12:~~{Construct overall  sample vector},~$\tilde{\br}_{\ell} \in \mathbb{R}^{2   | \mathcal{U}|}$
\State \textbf{{State update based on joint channel sounding}}
\State 13:~~{Compute overall combiner$, \tilde{\bZ}_{\ell}=\tilde{\bZ}_{\ell}^{u_1} \oplus \cdots \oplus  \tilde{\bZ}_{\ell}^{u_{|\mathcal{U}|}}$}
\State 14:~~{Design Kalman gain matrix,}
\State ~~~~~~$\tilde{\bK}_{\ell}=\hat{\bQ}_{\ell|\ell-1}   \big(( \bP_{\ell}^{\frac{1}{2}} \otimes \bI_2 )\tilde{\bZ}_{\ell}\tilde{\bD}_{\ell|\ell-1}\big)^T    \big( (\bP_{\ell}^{\frac{1}{2}} \otimes \bI_2 )\tilde{\bZ}_{\ell}$
\State ~~~~~~~~~~~~~~~$\tilde{\bD}_{\ell|\ell-1}\hat{\bQ}_{\ell|\ell-1}\big(( \bP_{\ell}^{\frac{1}{2}} \otimes \bI_2 )\tilde{\bZ}_{\ell}\tilde{\bD}_{\ell|\ell-1}\big)^T+{\bI_{2|\mathcal{U}|}}/{2}  \big)^{-1}$
\State 15:~~{Update state vector,}
\State ~~~~~~$\hat{\bt}_{\ell}=\hat{\bt}_{\ell|\ell-1}+\tilde{\bK}_{\ell}  ({\tilde{\br}_{\ell} - (\bP_{\ell}^{\frac{1}{2}} \otimes \bI_2 ) \tilde{\bZ}_{\ell} \hat{\bh}_{\ell|\ell-1}})$
\State 16:~~{Update covariance matrix,}
\State ~~~~~~$\hat{\bQ}_{\ell}=(\bI_2- \tilde{\bK}_{\ell} (  \bP_{\ell}^{\frac{1}{2}} \otimes \bI_2 )\tilde{\bZ}_{\ell} \tilde{\bD}_{\ell|\ell-1})\hat{\bQ}_{\ell|\ell-1}$
  \end{algorithmic}
\end{algorithm}

{The RSUs near a  vehicle can obtain a sounding sample  because an omnidirectional antenna, mounted on the roof of vehicles, radiates sounding signals in all directions.
{It is necessary to develop a joint tracking system to improve the vehicle tracking performance by jointly considering sounding samples at multiple RSUs.
One possible approach is to exploit three  samples, i.e., $\{\tilde{\br}_{\ell}^{1} ,\tilde{\br}_{\ell}^{2}, \tilde{\br}_{\ell}^{3}\}$, obtained from neighboring  RSUs.
However, this full cooperative  tracking solution would impose a burden on the backhaul network because  three  samples must be shared between RSUs for every sampling period.}

{This study focuses on developing  joint vehicle tracking system to enhance  tracking performance  while minimizing the amount of  sample exchange between neighboring RSUs.
A closer look at the full cooperative joint tracking system is needed to define the reference of   vehicle tracking performance.
Section \ref{sec:hand_over_algorithm} introduces the SANR metric to quantify  the vehicle tracking performance in terms of the derivative of spatial frequency, which depends on the position variables.
Assuming the Kalman gain matrix combines multiple sounding samples optimally, the overall SANR can be defined by $\bar{\gamma} \doteq \sum_{u=1}^{3} \gamma^{u}_{\ell|\ell-1}$.
We call the overall SANR of the full cooperative tracking system as the reference SANR.}

The overall SANR value depends on the quality of sounding samples used for joint vehicle tracking.
We aim to select RSUs as minimal as possible to produce the overall SANR that is greater than the predefined performance threshold $\tau_{\textrm{th}}$, by allowing a small amount of sample-exchange.
The SANRs are sorted in descending order, such as ${\gamma}^{u_1}_{\ell|\ell-1} \geq {\gamma}^{u_2}_{\ell|\ell-1} \geq {\gamma}^{u_3}_{\ell|\ell-1}$.
If a certain RSU satisfies the predefined  performance threshold  $\frac{{\gamma}^{u_1}_{\ell|\ell-1}}{\bar{\gamma}} \geq \tau_{\textrm{th}} $, only single RSU will be selected for vehicle tracking as $\mathcal{U}=\{ u_1\}$.
On the other hand, if two RSUs are needed to satisfy the performance threshold  $\frac{{\gamma}^{{u}_{1}}_{\ell|\ell-1}}{\bar{\gamma}}  < \tau_{\textrm{th}}$ and $\frac{{\gamma}^{{u}_{1}}_{\ell|\ell-1}+{\gamma}^{{u}_{2}}_{\ell|\ell-1}}{\bar{\gamma}}  \geq \tau_{\textrm{th}}$, a  set of selected RSUs will be given by $\mathcal{U}=\{ u_1, u_2\}$.
If strict subsets of RSUs cannot satisfy the performance threshold, all the RSUs must participate for joint vehicle tracking, such that\footnote{
{In our network deployment scenarios, the full set of RSUs, i.e., $\mathcal{U}=\{1,2,3\}$,  wouldn't be used in the proposed joint tracking algorithm.}} $\mathcal{U}=\{1,2,3\}$.
The service areas for the proposed joint tracking system are predefined based on SANR.
Active RSUs for joint  tracking will  be chosen based on the predefined service areas in Fig. \ref{fig:hand_over_example_03}.
{The service areas for the SNR-based joint tracking system are  predefined by substituting the SNR metric for the SANR metric in Algorithm \ref{Al:01}, as depicted  in Fig. \ref{fig:hand_over_example_04}.
The service areas for both systems are designed to have the same surface area for joint  tracking.}

{For a given set of selected RSUs $\mathcal{U}=\{ u_1,\cdots, u_{|\mathcal{U}|}\}$, an input-output expression for the overall channel sounding process can be defined by
\begin{align*}
\tilde{\br}_{\ell}=[(\tilde{\br}_{\ell}^{u_1})^T ,\cdots, (\tilde{\br}_{\ell}^{u_{|\mathcal{U}|}})^T]^T= (\bP_{\ell}^{\frac{1}{2}} \otimes \bI_2 ) \tilde{\bZ}_{\ell} \tilde{\bh}_{\ell} + \tilde{\bn}_{\ell} \in \mathbb{R}^{2   | \mathcal{U}|},
\end{align*}
where  $\bP_{\ell}=\mathrm{diag}([\rho_{\ell}^{u_1},\cdots,\rho_{\ell}^{u_{|\mathcal{U}|}}])$ is a diagonal matrix for the set of average SNRs, $\tilde{\bZ}_{\ell}=\tilde{\bZ}_{\ell}^{u_1} \oplus \cdots \oplus  \tilde{\bZ}_{\ell}^{u_{|\mathcal{U}|}}$ is  a block diagonal matrix for the overall combiner, $\tilde{\bh}_{\ell}=[(\tilde{\bh}_{\ell}^1)^T,\cdots,(\tilde{\bh}_{\ell}^{u_{|\mathcal{U}|}})^T]^T$ is the overall channel vector, and $\tilde{\bn}_{\ell}=[(\tilde{\bn}_{\ell}^{u_1})^T,\cdots, (\tilde{\bn}_{\ell}^{u_{|\mathcal{U}|}})^T]^T \sim \mathcal{N}\big(\mathbf{0}_{2|\mathcal{U}|},\frac{\bI_{2|\mathcal{U}|}}{2}\big)$ is the overall noise vector.
The  state update process is  designed as
\begin{align}
\label{eq:joint_state_update}
\hat{\bt}_{\ell}=\hat{\bt}_{\ell|\ell-1}+\tilde{\bK}_{\ell}\big({\tilde{\br}_{\ell} - (\bP_{\ell}^{\frac{1}{2}} \otimes \bI_2 ) \tilde{\bZ}_{\ell} \hat{\bh}_{\ell|\ell-1}}\big),
\end{align}
where $\hat{\bh}_{\ell|\ell-1}=[(\tilde{\bh}(\hat{\psi}_{\ell|\ell-1}^{u_1}))^T,\cdots,(\tilde{\bh}(\hat{\psi}_{\ell|\ell-1}^{u_{|\mathcal{U}|}}))^T]^T$ is the overall predicted channel vector.
The Kalman gain matrix is defined {as}
\begin{align}
\nonumber
&\tilde{\bK}_{\ell}=\hat{\bQ}_{\ell|\ell-1}   \big(( \bP_{\ell}^{\frac{1}{2}} \otimes \bI_2 )\tilde{\bZ}_{\ell}\tilde{\bD}_{\ell|\ell-1}\big)^T    \big( (\bP_{\ell}^{\frac{1}{2}} \otimes \bI_2 )\tilde{\bZ}_{\ell}\tilde{\bD}_{\ell|\ell-1}
\\
\label{eq:joint_kalman}
&~~~~~~~~~~~\hat{\bQ}_{\ell|\ell-1}\big(( \bP_{\ell}^{\frac{1}{2}} \otimes \bI_2 )\tilde{\bZ}_{\ell}\tilde{\bD}_{\ell|\ell-1}\big)^T+{\bI_{2|\mathcal{U}|}}/{2}  \big)^{-1}.
\end{align}
Lastly, the covariance matrix is refined as}
\begin{align*}
\hat{\bQ}_{\ell}=(\bI_2- \tilde{\bK}_{\ell} (  \bP_{\ell}^{\frac{1}{2}} \otimes \bI_2 )\tilde{\bZ}_{\ell} \tilde{\bD}_{\ell|\ell-1})\hat{\bQ}_{\ell|\ell-1}.
\end{align*}

{A set of selected RSUs for joint tracking can be predefined because the overall SANR is  a function of the vehicle position variables,  $(\hat{x}_{\ell|\ell-1},y)$, and the network design  parameters, $(\mathrm{X},\mathrm{Y},h)$.}
With the predicted state vector and the covariance matrix, $(\hat{\bt}_{\ell|\ell-1},\hat{\bQ}_{\ell|\ell-1})$, each RSU can construct  all the variables in (\ref{eq:joint_state_update}) required for {a} joint  tracking process, except the overall  sample vector, $\tilde{\br}_{\ell}$.
{Based on the predefined service area for joint tracking, as depicted in Fig. \ref{fig:hand_over_example_03}, it is required to exchange sounding samples between RSUs {to construct} the overall sample vector in the joint state update process.}

\section{Simulation Results}
\label{sec:numerical}

\begin{figure}
\centering
\subfigure[Position tracking performance.]{\label{fig:sim_hand_over_01}\includegraphics[width=0.2325\textwidth]{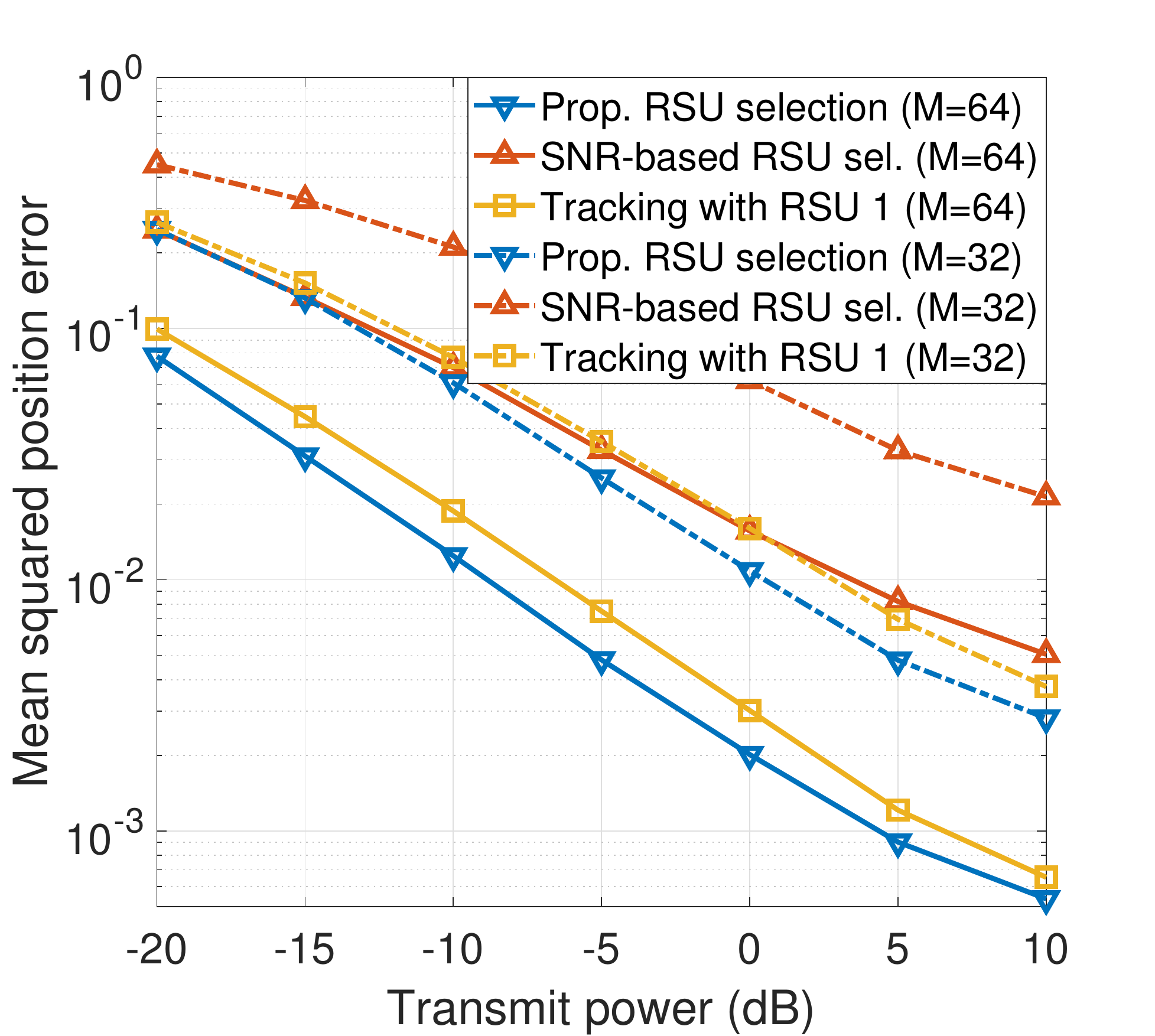}}
\subfigure[Velocity tracking performance.]{\label{fig:sim_hand_over_02}\includegraphics[width=0.2325\textwidth]{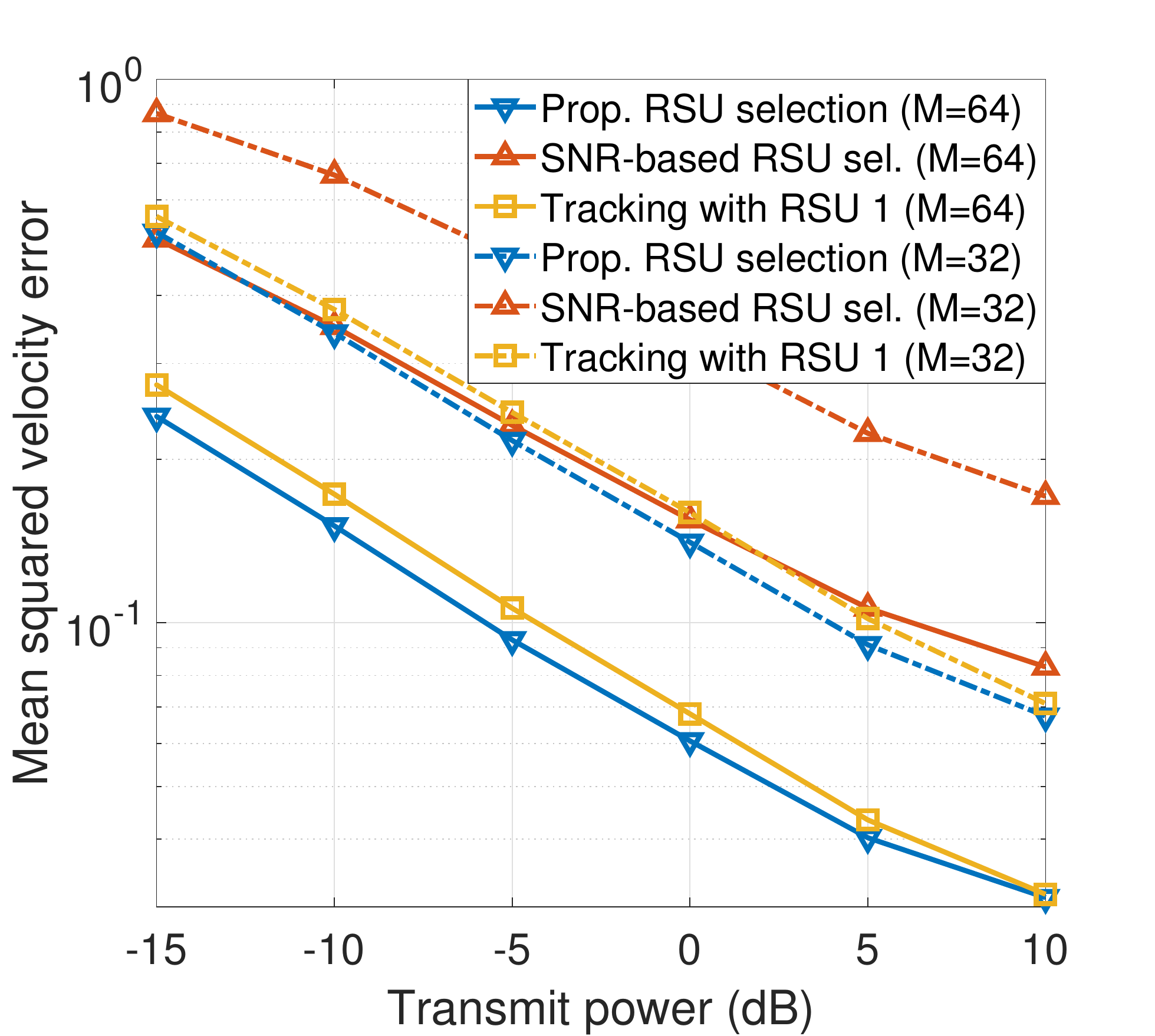}}
\caption{{Mean squared errors of RSU selection systems in \textit{Crossover Region B}.}}
\label{fig:hand_over}
\end{figure}

{The simulation} results are presented to evaluate the performance of the proposed vehicle tracking algorithms.
{This study generates} $10,000$ independent vehicle tracking scenarios.
The number of antennas at RSU is set to $M \in \{32,64\}$.
{The radio signals} at a center frequency $f_c=28$ GHz employing $20$ MHz bandwidth {are utilized.}
The noise power is then calculated as, $\sigma_{\mathrm{n}}^2=-174+10\log_{10}(20 \times 10^6)\simeq -101$ dBm.
The path-loss exponent is  $n=2$ and the sampling period for {the} uplink channel soundings is $T_{s}=10$ ms.
The channel vector consists of a line-of-sight and a non-line-of-sight radio path with Rician $K$ factor, $K=13$ dB.
Similar to \cite{Ref_Hyu22}, the error parameters in (\ref{eq:predict}) are set to $\{\sigma_{\omega},\sigma_{\alpha} \}=\big\{10^{-1.5},0.05\big(v_{0}\frac{10^3}{60^2}\big) \big\}$.

{Before evaluating the joint vehicle tracking system, we present the tracking performances of the proposed SANR-based RSU selection algorithms   within \textit{Crossover Region B} in Fig. \ref{fig:hand_over_example}.
The network design parameters are set to $(\mathrm{X},\mathrm{Y},h)=(75,31,7.5)$ m and  an initial velocity of vehicle is set to $v_0=60$ km/h ($16.67$ m/s).}
In Figs. \ref{fig:sim_hand_over_01} and \ref{fig:sim_hand_over_02},  {the}  position and velocity tracking performances are evaluated,  within {a} $2.5$ s duration, by using {the  mean squared errors, $\Upsilon_{x}=\mathrm{E}[|{x_{\ell}-\hat{x}_{\ell}}|^2]$ and $\Upsilon_{v}=\mathrm{E}[|{v_{\ell}-\hat{v}_{\ell}}|^2]$, respectively.}
{Fig. \ref{fig:hand_over} shows} that the proposed RSU selection algorithm using SANR gives better position and velocity tracking performances because  both an average SNR and angular variation owing to vehicle movements are jointly considered to evaluate {the} vehicle tracking performance.
The position estimation performance is better than the velocity estimation performance because the vehicle tracking algorithm is designed to use sounding samples  {written
in terms of the position variables.}

\begin{figure}
\centering
\subfigure[{Position tracking in \textit{{SANR-based RSU~1 Selection  Area}} with $(x_0,y)=(-75,3.25)$ m.}]{\label{fig:new_sim_joint_tracking_01}\includegraphics[width=0.4325\textwidth]{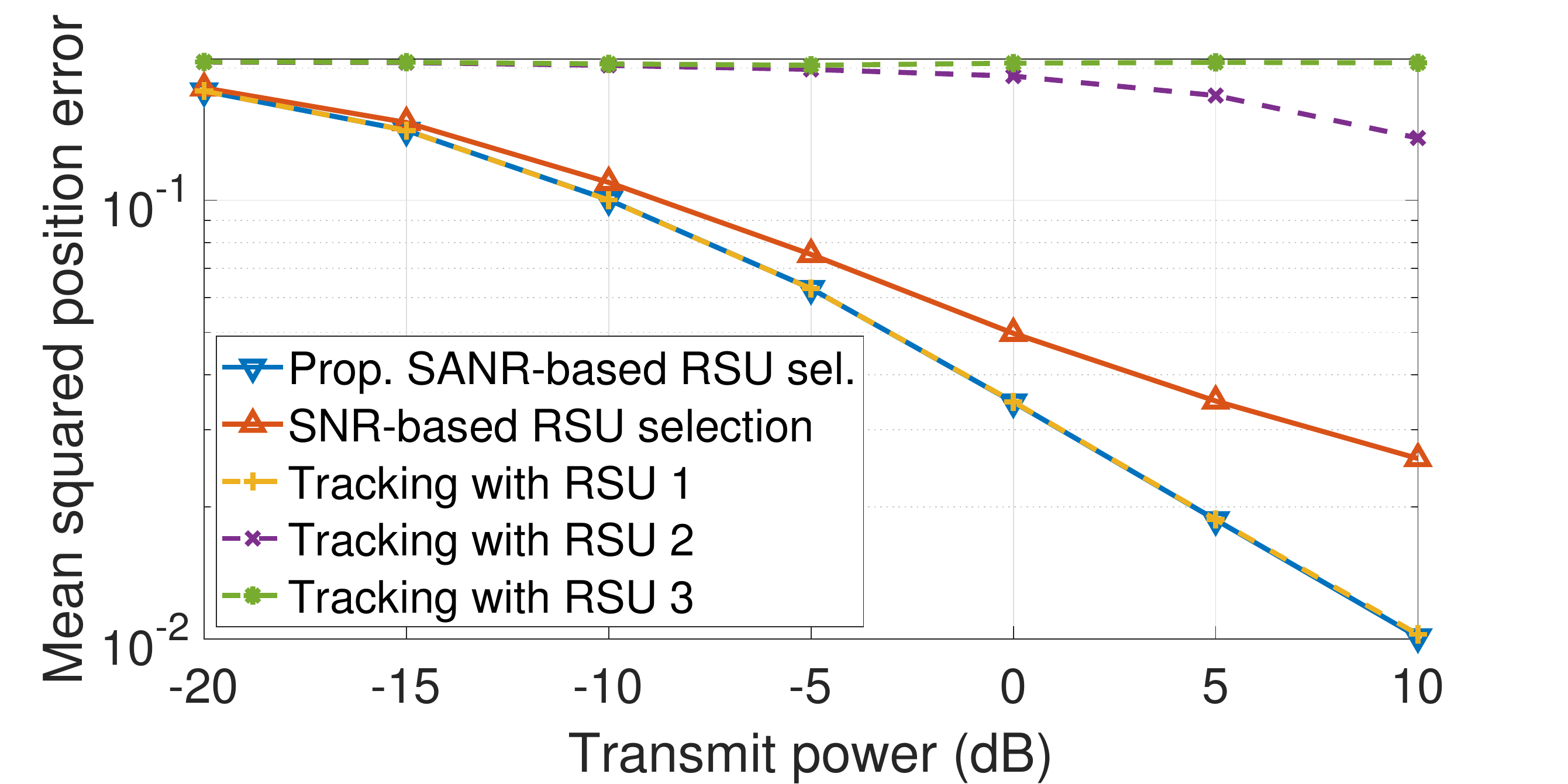}}
\subfigure[{Position tracking in \textit{{SANR-based RSU~2 Selection  Area}} with $(x_0,y)=(-80,24.25)$ m.}]{\label{fig:new_sim_joint_tracking_02}\includegraphics[width=0.4325\textwidth]{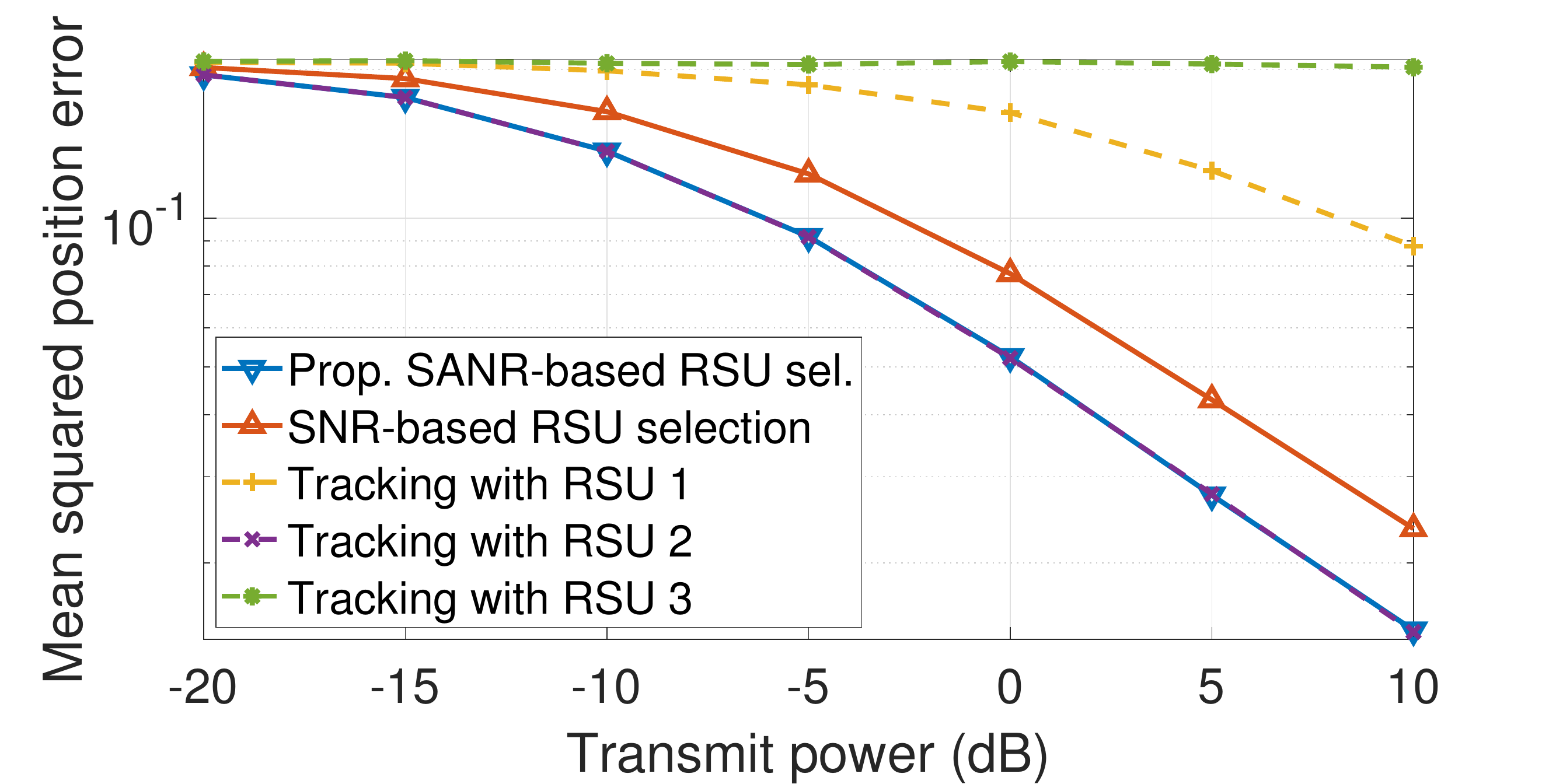}}
\caption{{Mean squared position errors of vehicle tracking systems with $M=32$ in \textit{{SANR-based  RSU~1 or 2 Selection Areas}}.}}
\label{fig:new_joint_tracking}
\end{figure}

{We evaluate position tracking performances by considering a single \textit{Service Area}  within {a} $1.5$ s duration.
We consider two different vehicle deployment scenarios, $(x_0,y) \in \{(-75,3.25)~\textrm{m}, (-80,24.25)~\textrm{m}\}$,
in which RSUs are deployed with the network design parameters, $(\mathrm{X},\mathrm{Y},h)=(125,31,7.5)$ m.
In the proposed SANR-based RSU selection algorithm, the RSU $1$ will always be chosen based on the SANR metric within  {\textit{SANR-based RSU 1 Selection Area}}.
In Fig. \ref{fig:new_sim_joint_tracking_01}, it is shown that there are no performance gap between the proposed RSU selection algorithm and the vehicle tracking system exploiting only RSU $1$.
For the above mentioned reasons, in Fig. \ref{fig:new_sim_joint_tracking_02}, the proposed RSU selection system and the vehicle tracking system exploiting only RSU $2$ produce the same vehicle tracking performances because the RSU $2$ will always be chosen based on the SANR metric within   {\textit{SANR-based RSU 2 Selection Area}}.}

\begin{figure}
\centering
\subfigure[Position tracking performance.]{\label{fig:sim_joint_tracking_01}\includegraphics[width=0.4685\textwidth]{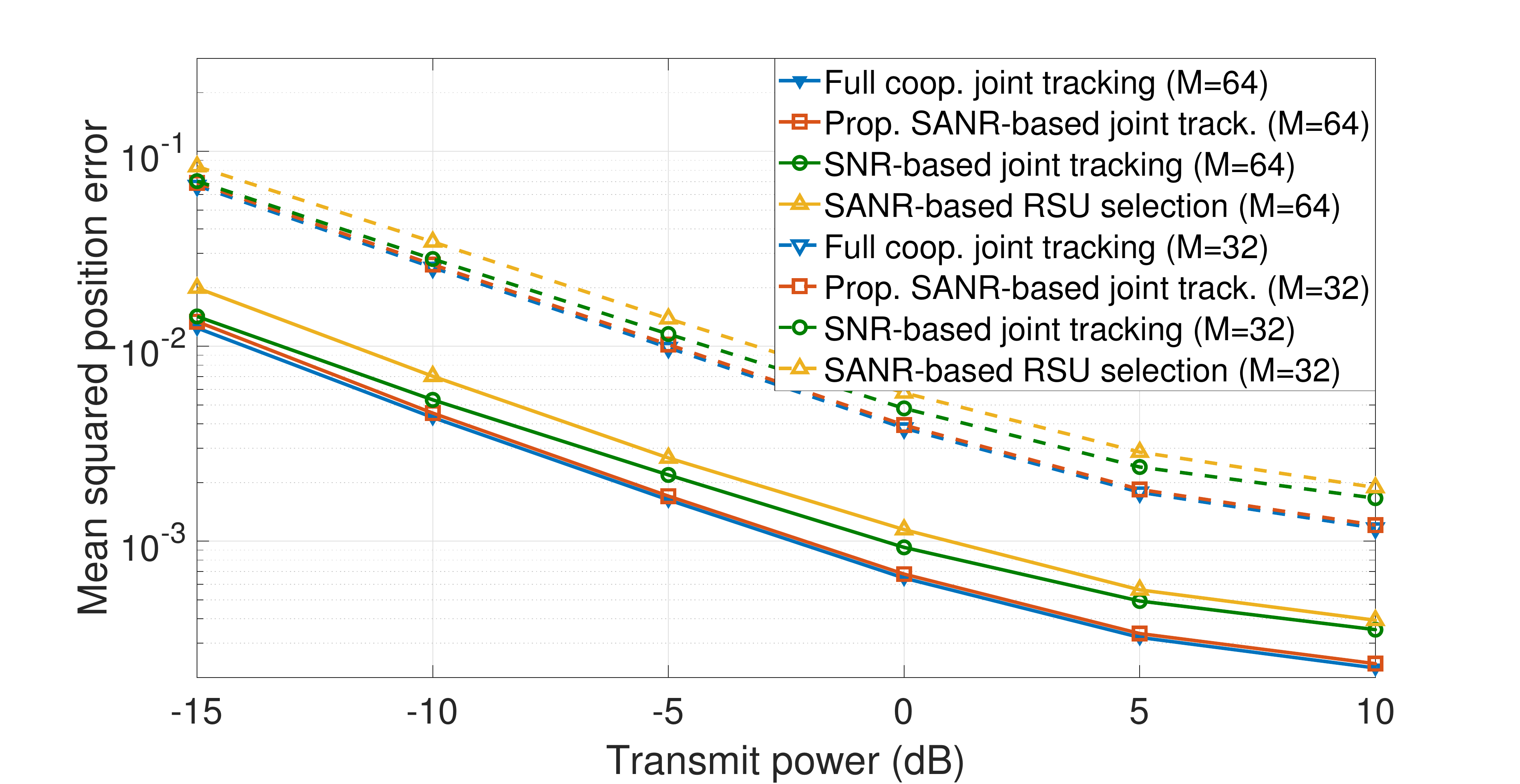}}
\subfigure[Velocity tracking performance.]{\label{fig:sim_joint_tracking_02}\includegraphics[width=0.4685\textwidth]{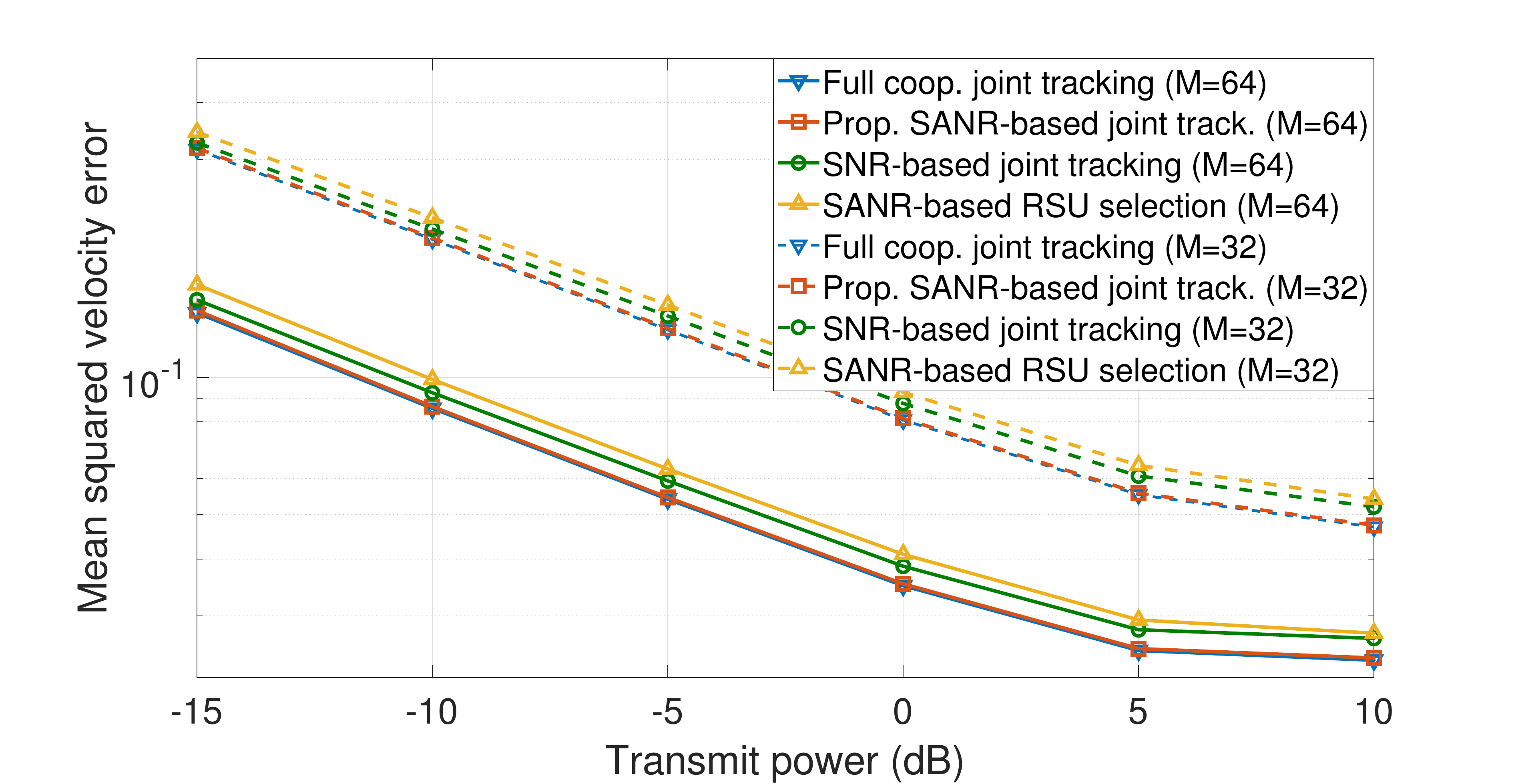}}
\caption{{Mean squared errors of vehicle tracking systems  in {\textit{{SANR-based RSU~1 \& 2 Selection Areas}}}.}}
\label{fig:joint_tracking}
\end{figure}

{Lastly, a vehicle tracking performance of the proposed joint tracking  system is evaluated   in {\textit{SANR-based RSU~1 \& 2 Selection Areas}}}.
The network design parameters are set to $(\mathrm{X},\mathrm{Y},h)=(75,31,7.5)$ m and an initial state vector is given by $\bt_{0}=[-60~\textrm{m},~60~\textrm{km/h}]^{T}$ with $y=3.25$ m.
A vehicle departs from the point  within  {\textit{SANR-based RSU~2 Selection Area}} and  arrives in {\textit{SANR-based RSU~1 Selection Area}} after driving for $2.5$ seconds.
{As shown in Fig. \ref{fig:joint_tracking}, the joint tracking systems provide better performance than previous tracking systems exploiting a single RSU because the changes in multiple beam directions are considered jointly for tracking vehicle movements.
The mean squared error decreases with increasing beamforming gain by exploiting more antennas at RSUs.
The proposed SANR-based joint tracking system with $\tau_{\textrm{th}}=0.98$ and the SNR-based joint tracking system with $\tau_{\textrm{th}}=0.662$  exploit $1.5$ sounding samples for joint vehicle tracking.
On the other hand, in the full cooperative tracking system, a vehicle tracking is performed by using sounding samples received from three RSUs.
It is verified that the  performance gap between the proposed joint tracking system  and  the full cooperative tracking system is negligible, although the proposed system exploits far less sounding  samples.
Furthermore, the proposed SANR-based joint tracking system outperforms the  SNR-based joint tracking  system that exploits similar amounts of samples.
Numerical results verify that the proposed SANR-based system enhances the vehicle tracking performance while minimizing the amount of sounding sample exchange.}

\section{Conclusion}
\label{sec:con}
We developed the  joint vehicle tracking and RSU selection algorithms by considering a massive RSU deployment scenario in V2I networks.
{The vehicle tracking performance was analyzed as a function of the} angular derivative of the dominant radio path in a spatial frequency domain and the  SNR.
Based on the derived metric, we developed an RSU selection algorithm that can maximize the vehicle tracking performance {by considering the relative position between a vehicle and RSUs.}
Moreover, a joint vehicle tracking algorithm {was} developed to track vehicle movements more reliably {while minimizing the exchange of sounding samples between  RSUs.}
{The} proposed joint   tracking and RSU selection algorithms {outperformed} conventional SNR-based vehicle tracking systems.

\bibliographystyle{IEEEtran}
\bibliography{ref}

\end{document}